\documentclass{Interspeech2024}

\interspeechcameraready

\usepackage{hyperref}
\title{Clever Hans Effect Found in Automatic Detection of Alzheimer's Disease through Speech\vspace{-1.2em}\thanks{* corresponding author. This work is supported in part by National Social Science Foundation of China (23AYY012). This research was supported by the Supercomputing Center of the USTC.}}
\name[affiliation={1}]{Yin-Long}{Liu}
\name[affiliation={1}]{Rui}{Feng}
\name[affiliation={1,2*}]{Jiahong}{Yuan}
\name[affiliation={1,2}]{Zhen-Hua}{Ling}
\address{
  $^1$National Engineering Research Center of Speech and Language Information Processing, \\ University of Science and Technology of China, Hefei, P. R. China\\
  $^2$Interdisciplinary Research Center for Linguistic Sciences, \\ University of Science and Technology of China, Hefei, P. R. China}
\email{\{lyl2001, fengruimse\}@mail.ustc.edu.cn, \{jiahongyuan, zhling\}@ustc.edu.cn}
\usepackage{multirow}
\usepackage{makecell}
\usepackage{graphicx}
\usepackage{hyperref}
\usepackage[bottom]{footmisc} 
\begin{document}

\maketitle

% the abstract here must exactly match the abstract entered into the paper submission system
\begin{abstract} 
We uncover an underlying bias present in the audio recordings produced from the picture description task of the Pitt corpus, the largest publicly accessible database for Alzheimer's Disease (AD) detection research. Even by solely utilizing the silent segments of these audio recordings, we achieve nearly 100\% accuracy in AD detection. However, employing the same methods to other datasets and preprocessed Pitt recordings results in typical levels (approximately 80\%) of AD detection accuracy. These results demonstrate a Clever Hans effect in AD detection on the Pitt corpus. Our findings emphasize the crucial importance of maintaining vigilance regarding inherent biases in datasets utilized for training deep learning models, and highlight the necessity for a better understanding of the models' performance. 
 
\end{abstract}
\vspace{-1mm}
\section{Introduction}
\vspace{-1mm}
Alzheimer's Disease (AD), the most common cause of dementia, is a neurodegenerative disease that worsens over time and causes irreversible damage to the brain, manifested by a persistent deterioration of an individual's cognitive and functional abilities, including language, memory, attention, and executive function \cite{nestor2004advances}. 

In recent years, researchers have achieved promising results in utilizing deep-learning models and an end-to-end approach for the automatic detection of AD through speech. However, the robustness of these models have not been thoroughly tested, primarily due to the scarcity of large and diverse datasets. 

In this paper, we reveal an inherent bias present in the audio recordings produced from the picture description task of the Pitt corpus from DementiaBank, the largest publicly accessible database for AD detection research. Remarkably, even by  exclusively utilizing the silent segments of these audio recordings, we achieve nearly 100\% accuracy in AD detection. As far as we are aware, this bias has not been reported in the literature. 
% Therefore, any studies involving the acoustic features of the original speech recordings from this dataset for AD detection will be affected by this bias.

We present this finding to draw researchers' attention to the impact of bias in the dataset, and advocate for more effort in studying the robustness and explainability of deep-learning models in automatic AD detection. 
\vspace{-1mm}
\section{Related work}
\vspace{-1mm}
\subsection{The Pitt corpus}
\begin{figure}[t]
    \centering
    \includegraphics[width=0.75\linewidth]{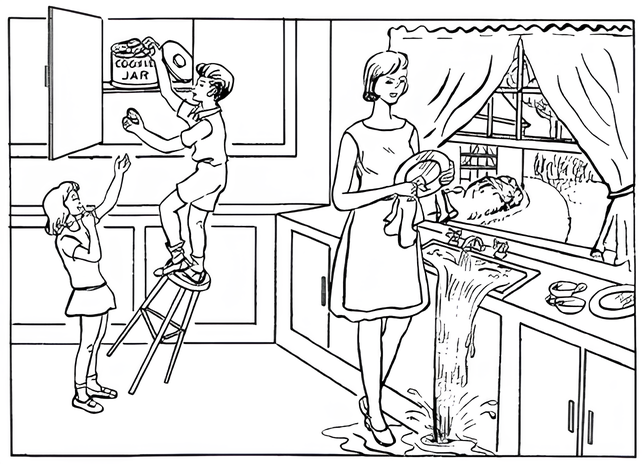}
    \caption{The picture of “Cookie Theft”, adopted from Boston
Diagnostic Aphasia Examination.}
    \label{fig:Cookie Theft}
     \vspace{-0.7cm}
\end{figure}

\vspace{-1mm}
% \textit{[Add a brief introduction to the Pitt corpus here, including the size of the data, and the noise in the recordings,noise is what?]} 
The Pitt corpus \cite{becker1994natural} is a widely used subset of DementiaBank. It was collected over a longitudinal period, encompassing 104 elderly controls, 208 individuals with probable or possible AD, and 85 participants with unknown diagnoses. Responses to four language tasks were recorded, including one task of describing the content of the Cookie Theft picture for all participants, which was originally designed for the Boston Diagnostic Aphasia Examination \cite{goodglass1983boston} (as shown in Figure~\ref{fig:Cookie Theft}), and three tasks of verbal fluency, sentence construction and story recall for AD participants only. For picture description task, there are 306 AD speech samples, 243 Healthy Controls (HC) speech samples, and due to the interference of the recording environment, these speech samples contain noise.
\vspace{-1mm}
\subsection{Automatic detection of AD through 
speech}
\vspace{-1mm}
Many researchers have utilized the Pitt corpus for AD detection studies. As of now, according to Google Scholar, this dataset has been cited in 543 papers, and the citation count is increasing year by year. 

Table~\ref{tab:previous studies} presents some results of previous studies using features extracted from the original, non-denoised recordings of the Pitt corpus. For example, Han et al. \cite{han2022automatic} utilized a miniature version of the Xception network \cite{chollet2017xception} to develop a deep learning model classifying AD and HC based on original speech selected from the Pitt corpus using log-mel spectrogram features, achieving an accuracy of 94.2\%. 
\begin{table}[b]
\vspace{-0.6cm}
\setlength{\belowdisplayskip}{0pt}
  \caption{Some results of previous studies using features extracted from the original, non-denoised recordings of the Pitt corpus.}
  \label{tab:previous studies}
  \centering
  \setlength{\tabcolsep}{1mm}{
  \begin{tabular}{l@{}cc} 
    \toprule
    \textbf{References}&\multicolumn{1}{c}{\textbf{Speech from Pitt}} & \textbf{Results(\%)}\\ 
    \midrule
 Han et al. \cite{han2022automatic}& 217 AD, 242 HC&94.2 Accuracy\\ 
 Ammar et al. \cite{ammar2021evaluation}& 43 AD, 43 HC&91  F-measure\\ 
 Zargarbashi et al. \cite{zargarbashi2019multi}& 255 AD, 233 HC&83.6 Accuracy\\ 
 Fraser et al. \cite{fraser2016linguistic}& 240 AD, 233 HC&81.92 Accuracy\\ 
 % Pan etal.\cite{pan2020improving}& 255 AD, 222 HC&78.34\% F-measure\\ 
 % Warnita etal.\cite{warnita2018detecting}& 255 AD, 233 HC&73.6\% Accuracy\\ 
 %    Krstev etal.\cite{krstev2022multimodal}&257 AD, 242 HC&73\% Accuracy\\
    \bottomrule
  \end{tabular}}
\end{table}

The Pitt corpus has been used for AD detection challenges at international conferences, including Interspeech 2020 (ADReSS) \cite{luz2020alzheimer} and Interspeech 2021 (ADReSSo) \cite{luz2021detecting}. In these challenges, the original recordings of the Pitt corpus underwent denoising and normalization process to create training and test data. Table~\ref{tab:previous studies of Challenge} presents the accuracy of previous studies using only audio recordings from challenge datasets for AD detection. As we can see, compared to the original, non-denoised data, the AD detection accuracy on the preprocessed data from these challenges was notably lower.  
\begin{table}[t]
 \caption{The accuracy of previous studies using only audio recordings from challenge datasets for AD detection.}
  \label{tab:previous studies of Challenge}
    \centering
    \begin{tabular}{ccc} 
    \toprule
         \textbf{References}&  \textbf{Dataset}& \textbf{Accuracy(\%)}\\ 
         \midrule
         Mei et al. \cite{mei2022detecting}&  ADReSS & 79.2\\ 
         Koo et al. \cite{koo2020exploiting}& ADReSS  & 72.9\\ 
        Gauder et al. \cite{gauder2021alzheimer}& ADReSSo &  78.9 \\
         Chen et al. 
         \cite{chen2021automatic}& ADReSSo & 77.1 \\ 
         \bottomrule
    \end{tabular}
\vspace{-0.7cm}
\end{table}

% \textit{[Add an summary of the challenge results, including methods and accuracies]}

The performance gap between using the original and preprocessed Pitt corpus may be attributed to two potential reasons. One is the improvement of the model's capability, and the other is the impact of spurious features present in the original speech data, i.e., the Clever Hans effect.
\vspace{-1mm}
\subsection{The clever hans effect}
\vspace{-1mm}
During the optimization process, models may exploit spurious correlations in the training data, resulting in seemingly high-performance metrics, a phenomenon known as the Clever Hans effect \cite{pfungst1911clever}. The Clever Hans refers to a horse that was believed to perform arithmetic and other intellectual activities. Subsequent investigations revealed that the horse did not actually execute these intellectual tasks. Instead, it responded to involuntary cues in the body language of its human trainer, of which the human trainer was entirely unaware. 

The Clever Hans effect is frequently observed in the context of supervised classifiers. Several machine learning problems have illustrated the effect. Arjovsky et al. \cite{arjovsky2019invariant} trained a convolutional neural network designed to classify camels and cows. 
% The majority of cow images in the training set were taken in green pastures, while most camel images were captured on beaches. However, the model failed to correctly classify straightforward examples of cow images taken on the beach. 
After experimental analysis, it was discovered that the neural network had successfully minimized its training error through a simple cheat: categorizing green landscapes as cows and beige landscapes as camels. 
Borah et al. \cite{borah2023measuring} mentioned Clever Hans behavior in high-performance neural translationese classifiers, where BERT-based
classifiers capitalize on spurious correlations, in particular topic information, between data and target classification labels, rather than genuine translationese signals. 
Chettri et al. \cite{chettri2018analysing} proposed that any visible pattern difference, such as the distribution of silence, between bonafide and spoof classes can introduce biases in voice spoofing detection, consequently influencing model decisions. Similar effect has also been demonstrated in medical contexts. Wallis et al. \cite{wallis2022clever} exposed an underlying bias in a commonly used publicly available brain tumour MRI dataset, and proposed that this is due to implicit radiologist input in the selection of the 2D slices. 
In the KDD CUP breast cancer identification challenge, Perlich etal. \cite{perlich2008breast} found that the patient IDs (which had not been removed from the data) were highly correlated with the malignancy of the patients’ tumours. 
Several recent studies \cite{degrave2021ai,lopez2021current} have revealed biases in datasets designed for COVID-19 identification from X-ray images, which stem from the inclusion of positive and negative images obtained from distinct sources. 
The biases described in the above-mentioned studies can lead machine learning models to take a ``shortcut" and address a significantly easier task. 
\vspace{-1mm}
\section{Unveiling the Clever Hans effect}
\vspace{-1mm}
\subsection{Data}
\vspace{-1mm}
In addition to employing speech recordings from the Pitt corpus, we also utilized Mandarin speech samples from iFLYTEK, as well as speech samples from ADReSS and ADReSSo, for comparative analysis.
\vspace{-1mm}
\subsubsection{Data from the original Pitt corpus}
\vspace{-1mm}
We selected 255 speech samples from 168 probable or possible participants and 242 speech samples from 99 HC participants. In addition to the Pitt corpus original (\textit{Pco}, original speech) dataset, based on the timestamp information in manual transcripts of the corresponding speech recordings, we also derived these three datasets: Pitt corpus subject (\textit{Pcsu}, containing only subject speech), Pitt corpus silence (\textit{Pcsi}, containing only silent speech), and Pitt corpus interviewer (\textit{Pci}, containing only interviewer speech). Given that a certain participant may have multiple corresponding speech samples, we randomly selected one of them to ensure that speech samples from a specific participant do not simultaneously appear in both the training and test datasets.
\vspace{-1mm}
\subsubsection{Data from a Mandarin AD dataset}
\vspace{-1mm}
The Mandarin data utilized in this paper is sourced from iFLYTEK, where subjects were recruited from the Department of Neurology
and the Department of Memory Clinic of Shanghai Tongji Hospital \cite{liu2019dementia,guo2020text,sheng2022dementia} and were instructed to undertake the same picture description task mentioned earlier. We selected 120 AD speech samples and 173 HC speech samples. Similar to Section 3.1.1, we obtained four datasets: Mandarin original (\textit{Mo}), Mandarin subject (\textit{Msu}), Mandarin silence (\textit{Msi}), and Mandarin interviewer (\textit{Mi}). Each speech sample corresponds to a unique participant.
\vspace{-1mm}
\subsubsection{Data from the ADReSS and ADReSSo challenges}
\vspace{-1mm}
ADReSS and ADReSSo challenges were hosted by Interspeech 2020 \cite{luz2020alzheimer} and Interspeech 2021 \cite{luz2021detecting} conferences respectively. Both challenge datasets are in English and underwent acoustic enhancement through noise removal and audio volume normalization. The ADReSS dataset contains 78 AD speech samples and 78 HC speech samples respectively. The ADReSSo dataset contains 122 AD speech samples and 115 HC speech samples. In addition to the ADReSS original (\textit{Ao}) and ADReSSo original (\textit{Aoo}) datasets, we also derived ADReSS silence (\textit{As}) and ADReSSo silence (\textit{Aos}) datasets using the pyannote\footnote{https://github.com/pyannote/pyannote-audio} voice activity detection (VAD) tools \cite{Bredin23}, since the
timestamp information provided by these two challenge datasets does not include silent intervals.
\vspace{-1mm}
\subsection{Fine-tuning wav2vec 2.0 for AD detecion}
\vspace{-1mm}
Wav2vec 2.0 is a framework for self-supervised learning of speech representations using contrastive loss \cite{baevski2020wav2vec}. 
% Building upon wav2vec 2.0, Facebook released wav2vec 2.0 XLSR \cite{conneau2020unsupervised}, which learns cross-lingual speech representations by pretraining a single model from the raw waveform of speech in multiple languages and experiments show that cross-lingual pretraining significantly outperforms monolingual pretraining. 
In previous work \cite{mei2023ustc}, we had demonstrated the effectiveness of fine-tuning wav2vec 2.0 for AD detection. In this paper, we fine-tuned the wav2vec 2.0 models  ``facebook/wav2vec2-large-xlsr-53" and ``wbbbbb/wav2vec2-large-chinese-zh-cn" with a sequence classification head on top (a linear layer with the sigmoid activation function over the average pooled output) on English and  Mandarin speech data respectively. The models are available in the HuggingFace’s Transformers library\footnote{https://huggingface.co/facebook/wav2vec2-large-xlsr-53}\footnote{https://huggingface.co/wbbbbb/wav2vec2-large-chinese-zh-cn}. 
We used 5-fold cross-validation to evaluate the models.

For fine-tuning the wav2vec 2.0 models, we set the batch size to 1, the gradient accumulation steps to 4, the number of training epochs to 15, the learning rate to $3\times10^{-5}$, the warmup ratio to 0.1, and the loss function was cross-entropy. we employed the \textit{Transformers.Trainer} as the optimizer. We converted the audio file format from mp3 to wav and converted the audio from stereo to mono, along with downsampling the audio data from 44.1kHz to 16kHz.
\vspace{-1mm}
\subsection{Results}
\vspace{-1mm}
The following three experiments progressively introduce how we unveil the Clever Hans effect.
% The results of fine-tuning wav2vec 2.0 on the aforementioned datasets are presented in Table~\ref{tab:unpreprocessed dataset}.

% \textit{[Rewrite the following two paragraphs to emphasize: 1. the accuracy of Pitt corpus is much higher than that of the other ones; 2. Using only silent segments achieve more than 98\% accuracy. You should also note here that we didn't use the durations of the silent segments, instead, all silent segments were concatenated into one piece as input for fine-tuning.]}
% \textit{[Delete the results from preprocessed data in Table 2. Make a separate table in the next section to list those results.]}
% \textit{[add interviewer section]}
% \textit{[You should probably delete the experiments and results of using the interviewers' speech only. People may argue that the interviewers speak differently to AD or non-AD subjects, so it is not surprising at all that using interviewers' speech only can achieve good results.]}
Initially, we treated the Pitt corpus as a normal dataset for AD detection research. We used \textit{Pco}, \textit{Pcsu}, \textit{Mo}, \textit{Msu}, \textit{Ao}, \textit{Aoo} to fine-tune the wav2vec 2.0 models respectively. Only speech recordings with a duration longer than 35 seconds were retained in the datasets, and only the first 35 seconds were used for fine-tuning. The results are shown in the corresponding rows of Table~\ref{tab:unpreprocessed dataset}. It can be seen that the accuracy of Pitt corpus is much higher than that of the other ones. Specifically, the classification accuracy of \textit{Pco} is much higher than that of the two challenge datasets \textit{Ao} (80.7\%), \textit{Aoo} (77.7\%) and the Mandarin dataset \textit{Mo} (81\%) and it is close to 100\% (97.2\%), a result that is worth pondering, since it should be comparable to the performance on \textit{Ao} and \textit{Aoo}. Considering the above results, we initially suspect that the speech recordings in the Pitt corpus are interfered by some factors.
\begin{table}[t]
    \caption{The AD detection accuracy of using different datasets to fine-tune wav2vec 2.0 models respectively.}
    % \textbf{Different subdataset}
    \label{tab:unpreprocessed dataset}
    \centering
    % \resizebox{\linewidth}{!}{
    \setlength{\tabcolsep}{0.8mm}{
    \begin{tabular}{lclc}
        \toprule
          \textbf{Dataset}&{\makecell[c]{\textbf{Different} \\ \textbf{subdataset}}}&{\makecell[c]{\textbf{Number of} \\ \textbf{training/test samples}}}&  \textbf{Accuacy{(\%)}}\\
         \midrule
          \multirow{3}*{Pitt corpus}&\textit{Pco} &{\makecell[c]{204/51}}& 97.2\\
          ~ &\textit{Pcsu} &{\makecell[c]{124/31}}&  90.3\\
          % ~ &Pcss&  96.5\%\verb|\|81.8\%\\
            ~ & \textit{Pcsi} &{\makecell[c]{208/52}}&98.9\\
            \midrule
          \multirow{3}*{Mandarin}&\textit{Mo} &{\makecell[c]{232/58}}&  81\\
          ~&\textit{Msu} &{\makecell[c]{136/34}}&  73.5\\
          % ~&Mss&  75.2\%\verb|\|75.7\%\\
 ~& \textit{Msi} &{\makecell[c]{52/13}}&57.3\\
 \midrule
          \multirow{2}*{ADReSS}&\textit{Ao} &{\makecell[c]{112/28}}&  80.7\\
~ & \textit{As} &{\makecell[c]{120/30}}&56.7\\
 \midrule
          \multirow{2}*{ADReSSo}&\textit{Aoo} &{\makecell[c]{176/44}}&  77.7\\
 ~& \textit{Aos} &{\makecell[c]{188/47}}&61.3\\
 \bottomrule
    \end{tabular}}
    % }
    \vspace{-0.4cm}
\end{table}

Next, for proving that the Pitt corpus indeed has problems, we conducted the second experiment. We fine-tuned the wav2vec 2.0 models using the first 85 seconds of each training sample in the two datasets, \textit{Pcsu} and \textit{Msu}, respectively, and conducted the test on the first 85 seconds of each speech sample from \textit{Pci} and \textit{Mi} datasets, respectively. The label of each speech sample in \textit{Pci} and \textit{Mi} is the same as the subject interviewed by the corresponding interviewer. The results are shown in Table~\ref{tab:subject train interviewer test}. The test performance on the \textit{Pci} dataset is an astonishing 83.1\%, a figure that seems unbelievable given that the model, trained exclusively on subjects' speech, theoretically lacks the ability to identify a subject's AD based solely on the interviewer's speech. The test performance on the \textit{Mi} dataset is relatively low, only 64.1\%, which is normal. Based on the above results, we can confirm that the Pitt corpus is definitely influenced by certain factors.
\begin{table}[t]
   \caption{The AD detection accuracy of the wav2vec 2.0 models fine-tuned with only subject speech on only subject speech or only interviewer speech respectively. }
   \label{tab:subject train interviewer test}
   \centering
   \begin{tabular}{ccc}
       \toprule
         \textbf{Training Set}&\textbf{Test Set}& \textbf{Accuacy{(\%)}}\\
        \midrule
         \multirow{2}*{\textit{Pcsu}}&\textit{Pcsu}& 98.1\\
         ~ &\textit{Pci}& 83.1\\
         \midrule
         \multirow{2}*{\textit{Msu}}&\textit{Msu}& 84.4\\
         ~ &\textit{Mi}& 64.1\\
         \bottomrule
   \end{tabular}
 \vspace{-0.7cm}
\end{table}

Then, in order to further analyze which specific factors interfere with the Pitt corpus , we attempted the third experiment. We used the first 35 seconds of each speech sample from \textit{Pcsi}, \textit{As}, \textit{Aos}, and \textit{Msi} to fine-tune the wav2vec 2.0 models, respectively. We didn't use the duration information of the silent segments, instead, all silent segments of each original speech sample were concatenated into one piece as input for fine-tuning. The label for the silence piece is the same as the corresponding subject. The results are shown in the corresponding rows of Table~\ref{tab:unpreprocessed dataset}. It can be seen that the performance on \textit{Pcsi} can reach 98.9\%, which is astonishing. On the contrary, the accuracy on \textit{As} (56.7\%) and \textit{Aos} (61.3\%) is much lower, and the accuracy on \textit{Msi} (57.3\%) is almost random guessing, as what we can expect. These results suggest that the audio recordings in Pitt corpus are interfered by environmental factors such as background noise. The models learned to capture these spurious features and correlations, leading to their high performance. 
\vspace{-1mm}
\section{Validating the Clever Hans effect}
\vspace{-1mm}
\subsection{Classification based on hand-crafted and wav2vec 2.0 features}
\vspace{-1mm}
To validate the bias of the recording environment in speech samples from the Pitt corpus, we employed the openSMILE toolkit \cite{eyben2010opensmile} to extract the ComParE 2016 features \cite{schuller2016interspeech} from speech recordings containing only silence, serving as our low-level acoustic features. ComParE 2016 is the largest feature set (6373 dimensions) in the toolkit and has been used for AD detection \cite{luz2021detecting,mei2023ustc,chen2021automatic}. We utilized XGBoost, GBDT, AdaBoost classifiers, and their majority voting, as provided by the scikit-learn package, to conduct 5-fold cross-validation on the aforementioned features for both the English silent speech dataset from the Pitt corpus and Mandarin silent speech dataset. 
Likewise, these classifiers have also been successfully applied to AD detection \cite{mei2023ustc, jeon2023early}.
% Because the manually annotated silent segments may include the voices of the subjects or interviewers, and considering how the attention mechanism of the wav2vec 2.0 models work, the features generated by these models tend to focus more on the speaker's voice. 
In addition, features from the last hidden layer of the wav2vec 2.0 model fine-tuned with the English silent speech dataset (\textit{Pcsi}) were also utilized for building classifiers. Both the original 1024-dimensional features and dimensionally reduced ones were explored. The reduction was achieved to 10 and 5 dimensions using Principal Component Analysis (PCA) from the scikit-learn package.
%In addition to that, the 1024-dimensional features from the last hidden layer of the wav2vec 2.0 models, which were fine-tuned with English silent speech dataset, were not only studied directly with machine learning classifiers but also dimensionally reduced using Principal Component Analysis (PCA) provided by the scikit-learn package. The dimension reduction was performed to 10 dimensions and 5 dimensions. After that, the dimensionally reduced features were trained and evaluated again with the aforementioned machine learning classifiers to conduct a more detailed analysis.
% Because manually annotated silent segments may contain the voice part of the subjects or interviewers, and considering the attention mechanism of the wav2vec 2.0 models, the features generated by the models will pay more attention to the speaker's voice part. Therefore, the 1024-dimensional features generated by the last hidden layer of wav2vec 2.0 models fine-tuned with silent speech were in addition to being studied directly with machine learning classifiers, they were also dimensionally reduced (10 dimensions and 5 dimensions) using Principal Component Analysis (PCA) provided by scikit-learn package, and then trained and evaluated again with the aforementioned machine learning classifiers to further analyze.
\begin{table}[b]
\vspace{-0.5cm}
    \caption{AD detection accuracy (\%) of  each machine learning classifier on the ComParE 2016 feature set of the \textit{Pcsi} and \textit{Msi} datasets.}
    \label{tab:ComParE 2016 train and test}
    \centering
    \begin{tabular}{ccccc}
        \toprule
         \textbf{Dataset} &  \textbf{XGBoost} &  \textbf{GBDT}&  \textbf{AdaBoost} & \textbf{Voting}\\
         \midrule
         \textit{Pcsi}&  83.8&  80.1&  87.6& 80.8\\ 
         \textit{Msi}&  49.3&  50.7&  55.5& 55.2\\ 
         \bottomrule
    \end{tabular}
\end{table}

 We used the aforementioned methods to study \textit{Pcsi} and \textit{Msi} to validate our hypothesis, by building classifiers on the ComParE 2016 feature set for these two datasets. Additionally, we explored the 1024-dimensional features generated by the fine-tuned wav2vec 2.0 model with the \textit{Pcsi} dataset, as well as the dimensionally reduced features (10 dimensions and 5 dimensions) of the fine-tuned wav2vec 2.0 model. The results are shown in Table~\ref{tab:ComParE 2016 train and test} and Table~\ref{tab:PCA train and test}. It can be observed that AdaBoost classifier can still achieve a high AD detection accuracy of 87.6\% based solely on \textit{Pcsi}'s low-level acoustic features ComParE 2016, while the performance of each classifier on \textit{Msi}'s ComParE 2016 is only about 50\%.  Moreover, whether it is on the original features (1024 dimensions) of the fine-tuned wav2vec 2.0 model with \textit{Pcsi} or on the features after dimensionality reduction, we can achieve an unbelievable nearly 100\% accuracy on the \textit{Pcsi} dataset, which contain only silences of the Pitt corpus. These results confirm an underlying bias in the speech recordings produced by the picture description task in the Pitt corpus, which is caused by interference from environmental factors such as background noise. Studies using acoustic features extracted from the original speech recordings of this dataset for AD detection will be affected by this bias.
 
\begin{table}[t]
    \caption{AD detection accuracy (\%) of  each machine learning classifier on the features generated by the fine-tuned wav2vec 2.0 model with \textit{Pcsi} dataset and the fine-tuned wav2vec 2.0 model's dimensionally reduced features.}
    \label{tab:PCA train and test}
    \centering
    \setlength{\tabcolsep}{1mm}{
    \begin{tabular}{ccccc}
        \toprule
         \textbf{Feature set}& \textbf{XGBoost}&  \textbf{GBDT}&  \textbf{AdaBoost}& \textbf{Voting}\\
         \midrule
         wav2vec 2.0 (1024)&  97.4&  98.9&  98.5& 98.9\\ 
         \midrule
         PCA(10) &  99.2&  97.7&  97.7& 97.7\\ 
         \midrule
         PCA(5)& 99.2& 98.1& 98.1&98.1\\
 \bottomrule
    \end{tabular}
    }
\vspace{-0.5cm}
\end{table}

\begin{figure}[b]
\vspace{-0.5cm}
    \centering
\includegraphics[width=0.75\linewidth]{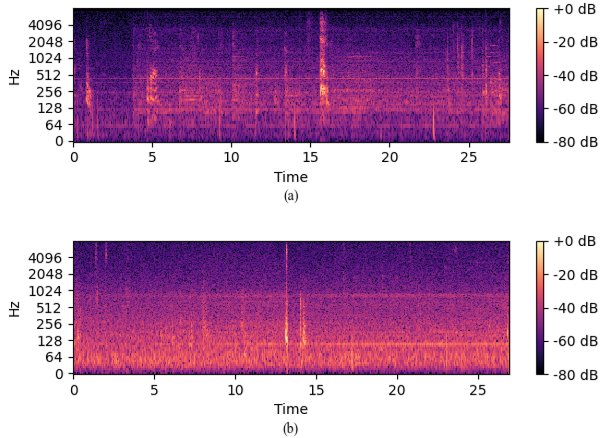}
    \caption{Spectrograms of (a) an AD speech sample and (b) an HC speech sample in \textit{Pcsi}.}
    \label{fig:Spectrogram of AD and HC speech samples in Pcs}
    % \vspace{-0.7cm}
\end{figure}

Figure~\ref{fig:Spectrogram of AD and HC speech samples in Pcs} depicts the spectrograms of two randomly selected silent segments (AD and HC) from \textit{Pcsi}. We can easily distinguish between AD and HC based on the two segments. This observation is consistent with the point made above.
\vspace{-1mm}
\subsection{Results from using preprocessed speech recordings}
\vspace{-1mm}
In order to mitigate the impact of the bias, we preprocessed the speech recordings before employing machine learning methods on the datasets. We utilized noisereduce package\footnote{https://github.com/timsainb/noisereduce} to reduce stationary noise \cite{sainburg2020finding}. The package relies on a method called ``spectral gating" which is a form of Noise Gate. It works by computing a spectrogram of a signal (and optionally a noise signal) and estimating a noise threshold (or gate) for each frequency band of that signal/noise. That threshold is used to compute a mask, which gates noise below the frequency-varying threshold. For stationary noise reduction, it should keep the estimated noise threshold at the same level across the whole signal. 
We also utilized the ``AudioSegment" and ``effects" methods of pydub package\footnote{https://github.com/jiaaro/pydub} for standard amplitude normalization, which scaled the whole audio to the max amplitude.

We utilized the methods described above to preprocess the \textit{Pco}, \textit{Pcsu}, \textit{Pcsi}, \textit{Mo}, \textit{Msu}, and \textit{Msi}, respectively. Subsequently, under the same experimental parameters configuration as the unpreprocessed datasets, we employed the preprocessed new speech datasets to fine-tune the wav2vec 2.0 models respectively. The results are presented in Table~\ref{tab:preprocessed dataset}. Comparing these outcomes with the corresponding results on original datasets, it is evident that the performance of the preprocessed \textit{Pcsi} dataset has significantly decreased (from 98.9\% to 63.1\%). This reduced performance is now comparable to that of \textit{Aos} (61.3\%). Moreover, the performance on the preprocessed \textit{Msi} remains at the level of random guessing (58.8\%). 
% Figure~\ref{fig:preprocessed_Pcsi_AD_HC_spectrogram} illustrates the spectrograms of two speech samples corresponding to Figure~\ref{fig:Spectrogram of AD and HC speech samples in Pcs} in the preprocessed Pcsi.  It is more intuitive to observe the effect of preprocessing, and now, the two spectrograms are similar.
Additionally, the performance on preprocessed \textit{Pco} and \textit{Pcsu} is reduced to what we consider a normal level. Specifically, the accuracy on \textit{Pco} decreases from 97.2\% to 82\%, which is essentially similar to the performance on \textit{Ao} (80.7\%). The results further confirm that speech recordings in the original Pitt corpus are affected by environmental interference, such as background noise. The performance of preprocessed \textit{Mo} and \textit{Msu} is basically equivalent to or slightly improved compared to that of the original data, indicating the preprocessing methods do not compromise the valuable information used to distinguish AD and HC and are effective for AD detection.

\begin{table}[t]
    \caption{The AD detection accuracy of using preprocessed Pitt and Mandarin datasets to fine-tune wav2vec 2.0 models. (The numbers in the parentheses correspond to results obtained using original Pitt and Mandarin datasets listed in Table~\ref{tab:unpreprocessed dataset}.)}
    % \textbf{Different subdataset}
    \label{tab:preprocessed dataset}
    \centering
    % \resizebox{\linewidth}{!}{
    \setlength{\tabcolsep}{0.8mm}{
    \begin{tabular}{lclc}
        \toprule
          \textbf{Dataset}&{\makecell[c]{\textbf{Different} \\ \textbf{subdataset}}}&{\makecell[c]{\textbf{Number of} \\ \textbf{training/test samples}}}&  \textbf{Accuacy{(\%)}}\\
         \midrule
          \multirow{3}*{Pitt corpus}&\textit{Pco} &{\makecell[c]{204/51}}& 82 (97.2)\\
          ~ &\textit{Pcsu} &{\makecell[c]{124/31}}&  77.4 (90.3)\\
          % ~ &Pcss&  96.5\%\verb|\|81.8\%\\
            ~ & \textit{Pcsi} &{\makecell[c]{208/52}}&63.1 (98.9)\\
            \midrule
          \multirow{3}*{Mandarin}&\textit{Mo} &{\makecell[c]{232/58}}&  80.7 (81)\\
          ~&\textit{Msu} &{\makecell[c]{136/34}}&  74.8 (73.5)\\
          % ~&Mss&  75.2\%\verb|\|75.7\%\\
 ~& \textit{Msi} &{\makecell[c]{52/13}}&58.8 (57.3)\\
 \bottomrule
    \end{tabular}}
    % }
    \vspace{-0.5cm}
\end{table}
\vspace{-1mm}
\section{Conclusions}
\vspace{-1mm}
In this paper, we expose an underlying bias present in the audio recordings produced from the picture description task of the Pitt corpus, a commonly used publicly available dataset for Alzheimer’s Disease detection. Even by solely leveraging the silent segments of these audio recordings, we can achieve nearly 100\% classification accuracy. Through experimental analysis, we propose that this bias is caused by background noise and other recording environment factors present in the original speech samples of the dataset. Subsequently, after preprocessing the speech samples with stationary noise removal and standard amplitude normalization, the experimental results demonstrate the alleviation of the bias, confirming the effectiveness of the data preprocessing methods. This study emphasizes the importance of understanding what the model has learned and calls for caution in the blind application of black-box automated models. We hope that researchers will pay attention to the potential dangers caused by spurious features in the data. In future work, we aim to delve into research on model interpretability, as well as explore cross-lingual Alzheimer’s Disease detection utilizing both English and Mandarin datasets.
\vspace{-2mm}
\bibliographystyle{IEEEtran}
\bibliography{mybib}
\end{document}